\pdfoutput=1
\documentclass[twocolumn,noversion,squeezed]{cdcarticle}

\def\MODE{2}

\usepackage{cite}                      
\usepackage{amsmath,amssymb,amsfonts}  
\usepackage{amsthm}                    
\usepackage[long,nocomma]{optiprob}    
\usepackage[utf8]{inputenc}				     
\usepackage{lmodern}
\usepackage[english]{babel}				     
\usepackage[hidelinks]{hyperref}	     
\usepackage{enumitem}					         
\usepackage{colonequals}               
\usepackage{textcomp}
\usepackage[T1]{fontenc}               
\usepackage{microtype}
\usepackage{scalerel,graphicx}         

\def\BibTeX{{\rm B\kern-.05em{\sc i\kern-.025em b}\kern-.08em
    T\kern-.1667em\lower.7ex\hbox{E}\kern-.125emX}}

\theoremstyle{definition}
\newcounter{lemcnt}
\newcounter{corcnt}
\newcounter{propcnt}
\newtheorem{thm}{Theorem}
\newtheorem{lem}[lemcnt]{Lemma}
\newtheorem{cor}[corcnt]{Corollary}
\newtheorem{prop}[propcnt]{Proposition}
\newtheorem*{rem}{Remark}
\newtheorem*{defn}{Definition}
\def\qed{\rule[0pt]{5pt}{5pt}\par\medskip}
\renewcommand{\qedhere}{\hfill ~\qed}

\setitemize{topsep=2pt,itemsep=-1pt}

\DeclareMathOperator*{\trace}{\mathrm{tr}}
\DeclareMathOperator*{\rank}{\mathrm{rank}}

\newcommand{\bmat}[1]{\begin{bmatrix}#1\end{bmatrix}}

\newcommand{\tp}{\mathsf{T}}
\newcommand{\defeq}{\colonequals}

\newcommand{\real}{\mathbb{R}}
\newcommand{\complex}{\mathbb{C}}
\newcommand{\symmetric}{\mathbb{S}}

\renewcommand{\i}{\ThisStyle{\setbox0=\hbox{$\SavedStyle\mathbb{i}$}\includegraphics[height=\ht0]{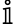}}}

\renewcommand{\L}{\mathcal{L}}
\newcommand{\I}{\mathcal{I}}
\newcommand{\M}{\mathcal{M}}

\renewcommand{\epsilon}{\varepsilon}

\newcommand{\lmin}{\lambda_\text{min}}

\renewcommand{\forall}{\text{ for all }}

\title{\LARGE \bf
Integral Quadratic Constraints:\\
Exact Convergence Rates and Worst-Case Trajectories}

\author{Bryan Van Scoy \and Laurent Lessard}

\note{}

\begin{document}

\maketitle

\begin{abstract}
We consider a linear time-invariant system in discrete time where the state and input signals satisfy a set of integral quadratic constraints (IQCs). Analogous to the autonomous linear systems case, we define a new notion of spectral radius that \textit{exactly} characterizes stability of this system. In particular, (i) when the spectral radius is less than one, we show that the system is asymptotically stable for all trajectories that satisfy the IQCs, and (ii) when the spectral radius is equal to one, we construct an unstable trajectory that satisfies the IQCs. Furthermore, we connect our new definition of the spectral radius to the existing literature on IQCs.
\end{abstract}

\if\MODE2%
{\let\thefootnote\relax\footnote{B.~Van~Scoy and L.~Lessard are with the Wisconsin Institute for Discovery at the University of Wisconsin--Madison, Madison, WI~53706, USA.\\
L.~Lessard is also with the Department of Electrical and Computer Engineering at the University of Wisconsin--Madison.\\
\texttt{\{vanscoy,laurent.lessard\}@wisc.edu}\\[1mm]
This material is based upon work supported by the National Science Foundation under Grants No. 1656951, 1710892, 1750162.}}%
\fi

\section{Introduction}

Consider the autonomous linear time-invariant system
\begin{align}\label{eq:autonomous}
x_{k+1} = A x_k, \quad k\geq 0, \quad x_0\in\real^n.
\end{align}
It is well-known that the asymptotic convergence rate of this system is characterized by the spectral radius,~$\rho(A)$. Specifically, the state converges to the origin for all initial conditions if and only if $\rho(A)<1$.  To show that this is the case, we can construct either a Lyapunov function or a non-convergent trajectory as follows.
\begin{itemize}
\item If $\rho(A)<1$, then the linear matrix inequality (LMI) $A^\tp P A - P\prec 0$ holds for some ${P\succ 0}$ in which case $V_k = x_k^\tp P x_k$ is a Lyapunov function that can be used to prove that the state converges to the origin~\cite{stein1952}.
\item If $\rho(A)=1$, then we can use the eigenvector of $A$ corresponding to the eigenvalue with unit magnitude to construct an initial condition such that the state does \textit{not} converge to the origin.
\end{itemize}

In this work, we generalize these results to the case where the system has inputs and satisfies a set of integral quadratic constraints (IQCs). We recover the results for autonomous systems as special cases, providing intuition for our contributions, which we summarize as follows.

\paragraph{Main contributions.}
We define a generalized notion of spectral radius for a linear time-invariant system whose state and input signals satisfy a set of IQCs and show that this corresponds to the \textit{exact} worst-case asymptotic convergence rate of the system.
\begin{itemize}
\item If the spectral radius is strictly less than one, we show that the system is asymptotically stable for all trajectories satisfying the IQCs (see Theorem~\ref{thm:stability}).
\item If the spectral radius is equal to one, we construct trajectories satisfying the IQCs such that the system is \textit{not} asymptotically stable (see Theorem~\ref{thm:example}).
\end{itemize}
In each case, we construct either a Lyapunov function or an unstable trajectory, both of which can give insight into how a specific system may perform in practice.

\paragraph{Literature review.}
Yakubovich introduced integral quadratic constraints in the 1970s to analyze systems with advanced nonlinearities; see~\cite{yakubovich1971,megretski1997}. Such constraints characterize a wide class of nonlinearities and uncertain quantities such as saturation, delay, sector-bounded nonlinearities, slope-restricted nonlinearities~\cite{ZamesFalb}, and time-varying quantities. Since then, IQCs have also been used to study linear time-varying~\cite{seiler2018}, delayed, and parameter-varying~\cite{pfifer2017} systems as well as first-order optimization algorithms~\cite{lessard2016}. Such systems are analyzed by replacing the troublesome component with a set of constraints that hold between its input and output (i.e., the IQCs).

IQCs can be formulated in both the frequency~\cite{megretski1997} and time~\cite{willems1972,seiler2015} domains, with the two approaches connected by Parseval's theorem and the Kalman--Yakubovich--Popov (KYP) lemma~\cite{rantzer1996}. Time-domain IQCs may be characterized as either \textit{hard} or \textit{soft} depending on whether the constraint holds for all finite times or only in the limit as time approaches infinity, respectively. In this paper, we consider multiple soft IQCs in the time domain.

Using an advanced version of the $S$-procedure on Hilbert spaces~\cite{megretski1993}, Megretski and Rantzer state that the frequency-domain IQC theorem~\cite[Theorem 1]{megretski1997} is both necessary and sufficient for robust stability with respect to multiple soft IQCs (see~\cite[Remark 4]{megretski1997}). Our results correspond to the time-domain version of this statement.

\paragraph{Notation.}
$\|\cdot\|$ denotes the $2$-norm. $\real^{n\times m}$ denotes the set of ${n\times m}$ real matrices, and $\symmetric^n$ denotes the set of ${n\times n}$ real symmetric matrices. The matrix inequality $A\succ B$ ($A\succeq B$) denotes that $A-B$ is positive (semi)definite.

\section{Problem setup}

Given matrices $A\in\real^{n\times n}$ and $B\in\real^{n\times m}$, we consider the discrete-time linear time-invariant system
\begin{align}\label{eq:system}
x_{k+1} &= A x_k + B u_k, \quad k\geq 0, \quad x_0\in\real^n
\end{align}
where $x_k\in\real^n$ is the state, $u_k\in\real^m$ is the input, and $x_0\in\real^n$ is the initial condition. To characterize the set of all possible system trajectories, we use integral quadratic constraints, which we define as follows. Our definition of IQCs is non-standard since the IQCs are static, but we choose this for ease of exposition; we refer the reader to Appendix~B for a comparison with the frequency-domain definition where the IQC itself also has dynamics.

\begin{defn}[IQC]
Given matrices ${M_i\in\symmetric^{n+m}}$ for ${i\in\I}$ where $\I$ is a finite index set, we say that the system~\eqref{eq:system} satisfies the \emph{integral quadratic constraints} (IQCs) defined by the set $\M\defeq\{M_i\}_{i\in\I}$ if and only if for all trajectories of the system there exists a finite scalar $\beta\in\real$ such that
\begin{align}\label{eq:soft}
\sum_{k=0}^{N-1} \bmat{x_k \\ u_k}^\tp\! M_i \bmat{x_k \\ u_k} \ge \beta
\end{align}
for all integers $N\ge 1$ and all $i\in\I$.
\end{defn}

Our goal is to characterize the asymptotic properties of the system~\eqref{eq:system} when the trajectories satisfy the IQCs~\eqref{eq:soft}. To do so, we use the following definitions.

\begin{defn}[Robust stability]
We say that system~\eqref{eq:system}~is
\begin{itemize}
\item \emph{robustly asymptotically stable} if and only if
\[
  \lim_{k\to\infty} \|x_k\| = 0
\]
\item \emph{robustly bounded} if and only if $\|x_k\|$ is uniformly bounded above for all $k\ge 0$
\end{itemize}
for all trajectories satisfying the IQCs~\eqref{eq:soft}.
\end{defn}

Just as the asymptotic properties of the autonomous linear system~\eqref{eq:autonomous} are characterized by the spectral radius of $A$, we will show that the asymptotic properties of the system~\eqref{eq:system} subject to the IQCs~\eqref{eq:soft} are characterized by the following spectral radius.

\begin{defn}[Spectral radius]
Given a tuple $(A,B,\M)$, we define the \emph{spectral radius}, denoted $\rho(A,B,\M)$, as the optimal value of the following optimization problem:
\begin{alignat}{3}
  &\makebox[\widthof{infimum}]{$\underset{\displaystyle \rho,\,P,\,\{\lambda_i\}}{\mathrm{infimum}}$} \quad && \mathrlap{\rho} \label{eq:rho} \\
  &\text{subject to} \quad &&& 0 &\succeq\bmat{A^\tp P A - \rho^2 P & A^\tp P B \\ B^\tp P A & B^\tp P B} + \sum_{i\in\I} \lambda_i\,M_i \nonumber \\
  &                        &&& \rho & > 0 \nonumber \\
  &                        &&& P    & \succ 0 \nonumber \\
  &                        &&& \lambda_i & \ge 0 \quad\forall i\in\I, \nonumber
\end{alignat}
where $\rho\in\real$, $P\in\symmetric^n$, and $\lambda_i\in\real$ for all $i\in\I$.
\end{defn}

The optimization problem~\eqref{eq:rho} is non-convex. However, we make the following observations:
\begin{itemize}
\item Determining whether there exists a feasible point for some fixed $\rho>0$ is a linear matrix inequality.
\item There exists a feasible point for any $\rho>\rho(A,B,\M)$.
\item There does \textit{not} exist a feasible point for any ${\rho<\rho(A,B,\M)}$. 
\end{itemize}
Therefore, we can efficiently compute the spectral radius by performing a bisection search over $\rho$, where we solve a linear matrix inequality at each iteration.

To simplify the notation, we define the discrete-time \emph{Lyapunov operator} $\L : \symmetric^n\to\symmetric^{n+m}$ as
\begin{subequations}\label{eq:lyap_operator}
\begin{align}
\L(P) &\defeq \bmat{A^\tp P A - P & A^\tp P B \\ B^\tp P A & B^\tp P B}
\end{align}
along with its adjoint operator $\L^* : \symmetric^{n+m}\to\symmetric^n$ given by
\begin{align}
\L^*(Q) \defeq \bmat{A\! & B} Q \bmat{A\! & B}^\tp - \bmat{I\! & 0} Q \bmat{I\! & 0}^\tp.
\end{align}
\end{subequations}
Note that for all $P\in\symmetric^n$ and $Q\in\symmetric^{n+m}$ we have
\begin{align*}
\langle Q,\L(P)\rangle = \langle \L^*(Q),P\rangle
\end{align*}
where $\langle A,B\rangle \defeq \trace(A^\tp B)$ is the Frobenius inner product.

\section{Robust stability}

\begin{thm}[Robust stability]\label{thm:stability}
Consider the system~\eqref{eq:system} subject to the IQCs~\eqref{eq:soft}, and let $\rho\defeq\rho(A,B,\M)$.
\begin{itemize}
\item[\textbf{(a)}] The system is robustly asymptotically stable if $\rho<1$.
\item[\textbf{(b)}] The system is robustly bounded if $\rho\le 1$ and the optimum in~\eqref{eq:rho} is attained.
\end{itemize}
\end{thm}

\begin{rem}
It may be the case that the optimum in~\eqref{eq:rho} is \textit{not} attained. To illustrate this, consider the following example with no inputs and no IQCs:
\begin{align*}
A = \bmat{1 & 1 \\ 0 & 1}, \qquad
B = 0\in\real^{2\times 0}, \qquad
\M \text{ empty}.
\end{align*}
The spectral radius is $\rho(A,B,\M) = \rho(A) = 1$, but there does not exist $P\succ 0$ such that $A^\tp P A - P\preceq 0$. Also, this system is \textit{not} bounded since the norm of the state grows unbounded with the initial condition $x_0 = \left[\begin{smallmatrix} 1 \\ 1 \end{smallmatrix}\right]$.
\end{rem}

\noindent\textbf{Proof of Theorem~\ref{thm:stability}.}
Let $(x_k,u_k)$ be a trajectory of the system~\eqref{eq:system} that satisfies the IQCs~\eqref{eq:soft}.\smallskip

\noindent\textbf{(a)} Suppose $\rho<1$. Then there exist $P\succ 0$ and $\lambda_i\ge 0$ for all $i\in\I$ such that the linear matrix inequality
\begin{align}\label{eq:LMI1}
\L(P) + \sum_{i\in\I} \lambda_i\,M_i \preceq -\bmat{I_n & 0 \\ 0 & 0}
\end{align}
holds (even if the optimum is not attained) since~\eqref{eq:rho} is homogeneous in $(P,\lambda_i)$, where the Lyapunov operator~$\L$ is defined in~\eqref{eq:lyap_operator}. To prove stability, we use the Lyapunov function
\begin{align}\label{eq:V}
V_k \defeq x_k^\tp P x_k + \sum_{i\in\I} \lambda_i \sum_{j=0}^{k-1} \bmat{x_j \\ u_j}^\tp\! M_i \bmat{x_j \\ u_j},
\end{align}
which is uniformly bounded below since the IQCs are satisfied. Also, the difference $\Delta V_k \defeq V_{k+1} - V_k$ satisfies
\begin{align}\label{eq:dV}
\Delta V_k = \bmat{x_k \\ u_k}^\tp\! \Bigl(\L(P)+\sum_{i\in\I}\lambda_i\,M_i\Bigr) \bmat{x_k \\ u_k},
\end{align}
where $\Delta V_k \le -\|x_k\|^2$ since~\eqref{eq:LMI1} holds. The sequence $V_k$ is monotonically decreasing and bounded below, so it converges to a constant. Then $\Delta V_k$ converges to zero as $k\to\infty$, so by the squeeze theorem we have that $\|x_k\|$ also converges to zero as $k\to\infty$. Thus, the system is robustly asymptotically stable.\smallskip

\noindent\textbf{(b)} Now suppose $\rho\le 1$ and the optimum is attained. Then there exist $P\succ 0$ and $\lambda_i\ge 0$ for all $i\in\I$ that satisfy the linear matrix inequality
\begin{align*}
\L(P) + \sum_{i\in\I} \lambda_i\,M_i \preceq 0.
\end{align*}
Using the Lyapunov function~\eqref{eq:V}, we have from~\eqref{eq:dV} that $\Delta V_k\le 0$, so $V_k\le V_0$ for all $k\ge 0$. Then using the bound $\lmin(P)\,\|x_k\|^2 \le x_k^\tp P x_k$, where $\lmin(P)$ denotes the minimum eigenvalue of $P$, we have
\begin{align*}
&\lmin(P)\ \limsup_{k\to\infty}\ \|x_k\|^2 \\
  &\qquad\le \limsup_{k\to\infty}\ x_k^\tp P x_k \\
  &\qquad= \limsup_{k\to\infty}\ \Bigl(V_k - \sum_{i\in\I} \lambda_i \sum_{j=0}^{k-1} \bmat{x_j \\ u_j}^\tp\! M_i \bmat{x_j \\ u_j}\Bigr) \\
  &\qquad\le V_0 - \beta \sum_{i\in\I} \lambda_i \\
  &\qquad < \infty,
\end{align*}
where $\beta$ is the lower bound on the IQCs defined in~\eqref{eq:soft}. Then since $\lmin(P)>0$, the limit superior of $\|x_k\|$ as ${k\to\infty}$ is finite, which implies $\|x_k\|$ is uniformly bounded above, so the system is robustly bounded. \qedhere

\begin{cor}[Robust exponential stability]
Suppose the optimum in~\eqref{eq:rho} is attained, and let $\rho\defeq\rho(A,B,\M)$. Then the system~\eqref{eq:system} is robustly exponentially stable with rate~$\rho$ with respect to the $\rho$-weighted IQCs. In other words, for all trajectories of the system such that for some $\beta\in\real$ the inequality
\begin{align*}
\sum_{k=0}^{N-1} \rho^{-2k} \bmat{x_k \\ u_k}^\tp\! M_i \bmat{x_k \\ u_k} \ge \beta
\end{align*}
holds for all integers $N\ge 1$ and all ${i\in\I}$, there exists a constant $c>0$ such that
\begin{align*}
\|x_k\|\le c\,\rho^k \quad\forall k\ge 0.
\end{align*}
\end{cor}

\noindent\textbf{Proof.}
The proof follows from applying Theorem~\ref{thm:stability} to the $\rho$-weighted trajectory $(\rho^{-k} x_k,\rho^{-k} u_k)$. \qedhere

\section{Worst-case trajectories}

We now show that Theorem~\ref{thm:stability} is tight by constructing trajectories that satisfy the IQCs and are such that the system is \textit{not} asymptotically stable when the spectral radius is equal to one.

Suppose $\rho(A,B,\M)=1$ and~\eqref{eq:rho} attains its optimum. Then the Lyapunov function~\eqref{eq:V} is non-increasing so the system is robustly bounded from Theorem~\ref{thm:stability}. However, the system may also be asymptotically stable if there are no trajectories such that the Lyapunov function is constant for all iterations. This follows from LaSalle's invariance principle, which is used to prove asymptotic stability when the Lyapunov function does not strictly decrease. To avoid such situations, we require a technical condition to hold. Before stating this condition, we need the following lemma, which we prove along with the main result (Theorem~\ref{thm:example}) in Appendix~A.

\begin{lem}\label{lem:XUF}
Suppose that $\rho(A,B,\M)=1$ and $B$ is full column rank. Then for some $d\ge 1$ there exists a tuple
\begin{align*}
(X,U,F) \in \real^{n\times d} \times \real^{m\times d} \times \real^{d\times d}
\end{align*}
with $X$ nonzero and $F$ orthogonal such that
\begin{subequations}\label{eq:XUF}
\begin{align}\label{eq:XUF1}
AX + BU = XF
\end{align}
and
\begin{align}\label{eq:XUF2}
\trace\Bigl(\bmat{X \\ U}^\tp\! M_i \bmat{X \\ U}\Bigr) \ge 0 \quad\forall i\in\I.
\end{align}
\end{subequations}
\end{lem}

\noindent\textbf{Technical condition.}
For some $(X,U,F)$ in Lemma~\ref{lem:XUF}, there exists a vector $v\in\real^d$ not in the null space of $X$ such that
\begin{align}\label{eq:technical}
v^\tp \Biggl(\sum_{j=1}^r W_j W_j^* \bmat{X \\ U}^\tp\! M_i \bmat{X \\ U} W_j W_j^*\Biggr) v \ge 0
\end{align}
for all $i\in\I$, where $F = W D W^*$ with 
\begin{subequations}\label{eq:WD}
\begin{align}
W &= \bmat{W_1 & \ldots & W_r} \\
D &= \text{diag}(e^{\i\theta_1} I,\ldots,e^{\i\theta_r} I),
\end{align}
\end{subequations}
where $W$ is unitary, $\theta_j\in[0,2\pi)$ are distinct, $W$ and~$D$ are partitioned conformably, and $r$ is the number of distinct eigenvalues of $F$. In other words, the number of columns in $W_j$ is the size of the $j\textsuperscript{th}$ identity matrix in $D$ which is also the multiplicity of the eigenvalue $e^{\i\theta_j}$ of $F$.

\begin{thm}[Worst-case trajectory]\label{thm:example}
Suppose $B$ is full column rank, ${\rho(A,B,\M)=1}$, and there exists ${v\in\real^d}$ that satisfies the technical condition for some $(X,U,F)$ from Lemma~\ref{lem:XUF}. Then the trajectory
\begin{align}\label{eq:trajectory}
\bmat{x_k \\ u_k} = \bmat{X \\ U} F^k v, \quad k\ge 0
\end{align}
satisfies the dynamics~\eqref{eq:system} as well as the IQCs~\eqref{eq:soft}, and the system is \textit{not} asymptotically stable.
\end{thm}

\begin{rem}[Static state feedback]
If $X$ is full column rank, then the trajectories~\eqref{eq:trajectory} are equivalent to using the initial condition $x_0 = X v$ and static state feedback $u_k = K x_k$ with gain matrix $K = U X^\dagger$, where~$(\cdot)^\dagger$ denotes the Moore--Penrose pseudoinverse.
\end{rem}

\subsection{Comments on the technical condition}\label{sec:technical}

We first motivate the technical condition by providing a simple example for which the technical condition fails and the system is robustly asymptotically stable even though the spectral radius is equal to one.\medskip

\noindent\textbf{Example.} Consider the following example with one state, no inputs, and two IQCs:
\begin{align*}
A  = 1\in\real^{1\times 1}, \qquad
B  = 0\in\real^{1\times 0}, \qquad
\M = \{\pm 1\}.
\end{align*}
The spectral radius of this system is equal to one and is achieved by the solution $P=1$ and ${\lambda_i=0}$ for all $i\in\I$, so the system is robustly bounded from Theorem~\ref{thm:stability}. In this case, however, the only trajectory that satisfies the IQCs is the trivial trajectory, i.e., $x_k=0$ for all $k\ge 0$. To see this, note that the state remains the same for all iterations, i.e., $x_k=x_0$ for all $k\ge 0$. Then in order for the IQCs to be satisfied, the initial condition $x_0$ must satisfy $x_0^\tp M_i x_0\ge 0$ for all $i\in\I$. But this has only the trivial solution $x_0=0$. Therefore, the system is robustly asymptotically stable even though $\rho(A,B,\M)=1$.\smallskip

While this example shows that we cannot construct unstable trajectories for all systems with spectral radius equal to one, we can under the following condition.

\begin{prop}
If all the eigenvalues of $F$ are distinct, then the technical condition is satisfied with $v = \sum_{j=1}^r W_j$ where $W_j$ is defined in~\eqref{eq:WD}.
\end{prop}

Note that all of the $W_j$ are column vectors in this case so they can be summed. Also, $v$ is a real vector even though the matrix~$W$ is complex since the columns of~$W$ form complex conjugate pairs and therefore have a real sum. This condition is trivially satisfied when $d=1$ in Lemma~\ref{lem:XUF} since the matrix $F$ is then a scalar.

\begin{prop}
If Lemma~\ref{lem:XUF} is satisfied with $d=1$, then the technical condition is satisfied with $v=1$.
\end{prop}

Determining whether the technical condition holds is in general NP-hard. However, there are several approaches for approximating the solution. One such approach uses the positivstellensatz from real algebraic geometry. This approach provides a hierarchy of semidefinite programs whose solutions converge to the solution of the original problem~\cite{parillo2003}. As an alternative, we can formulate the problem as a rank-constrained semidefinite program and take its convex relaxation~\cite{recht2010}. This yields the following convex optimization problem:
\begin{mini*}
{V\in\symmetric^d}{\|V\|_*}{}{}
\addConstraint{V\succeq 0}{}
\addConstraint{\trace(V X^\tp X)=1}{}
\addConstraint{\sum_{j=1}^r \trace\Bigl(V\, W_j W_j^* \bmat{X \\ U}^\tp\! M_i \bmat{X \\ U} W_j W_j^*\Bigr)\ge 0}{}
\addConstraint{}{\hspace*{3.9cm}\quad\forall i\in\I},
\end{mini*}
where $\|\cdot\|_*$ denotes the nuclear norm. If this problem has a rank-one solution, then the technical condition is satisfied by $v\in\real^d$ where $V = v v^\tp$. If the solution is not rank one, another approach is to apply a rank-reduction algorithm to find a low-rank solution~\cite{lemon2016}, although this is not guaranteed to find the minimal rank solution.

\subsection{Other types of IQCs}

The IQCs in~\eqref{eq:soft} are referred to as \textit{soft} since they restrict the trajectories only in the limit as time goes to infinity (the sum~\eqref{eq:soft} is always uniformly lower bounded by some $\beta$ for any finite~$N$). Using such IQCs to model nonlinearities and uncertainties in a system may be conservative. To address this, we now discuss two scenarios in which we can construct a worst-case trajectory that satisfies a more restrictive class of IQCs.\medskip

\noindent\textbf{Hard IQC.} Suppose there is a single IQC and the worst-case trajectory~\eqref{eq:trajectory} is such that the sum~\eqref{eq:soft} attains its lower bound for some index $N_\star$, in other words,
\begin{align*}
N_\star \in \arg\min_{N\ge 1}\ \sum_{k=0}^{N-1} v^\tp (F^k)^\tp \bmat{X \\ U}^\tp\! M \bmat{X \\ U} F^k v.
\end{align*}
Then the shifted worst-case trajectory
\begin{align*}
\bmat{x_k \\ u_k} = \bmat{X \\ U} F^{(k+N_\star)} v, \quad k\ge 0
\end{align*}
satisfies the dynamics~\eqref{eq:system} and the \emph{hard} IQC defined by
\begin{align}\label{eq:hard}
\sum_{k=0}^{N-1} \bmat{x_k \\ u_k}^\tp\! M \bmat{x_k \\ u_k} \ge 0
\end{align}
for all $N\ge 1$, and the system is \textit{not} asymptotically stable. The hard IQC is equivalent to the soft IQC with $\beta=0$.\medskip

\noindent\textbf{Pointwise IQC.} Suppose the worst-case trajectory~\eqref{eq:trajectory} has $d=1$ in Lemma~\ref{lem:XUF}. Then the trajectory also satisfies the \emph{pointwise} IQCs defined by
\begin{align}\label{eq:pointwise}
\bmat{x_k \\ u_k}^\tp\! M_i \bmat{x_k \\ u_k} \ge 0
\end{align}
for all $k\ge 0$ and all $i\in\I$. This follows from~\eqref{eq:XUF2} since $X\in\real^n$ and $U\in\real^m$ are vectors in this case.

\section{Conclusion}
In this paper, we generalized the notion of spectral radius for a matrix to that of a discrete-time system whose input and output signals satisfy a set of soft IQCs. When the spectral radius is strictly less than one, we constructed a quadratic Lyapunov function to prove that the system is robustly asymptotically stable. When the spectral radius is equal to one, we constructed a worst-case trajectory, and this trajectory is often equivalent to static state feedback. In order to construct the trajectory, we required a technical condition to be satisfied, and we provided several cases under which this holds. It remains an open question as to whether this technical condition is also necessary for constructing such a worst-case trajectory.

\appendix 
\section*{Appendix A --- Proofs}\label{sec:proof}

We begin by showing that strong duality holds between the following primal-dual semidefinite program pair; see Chapter 5 of~\cite{boyd2004} for an overview on Lagrangian duality.

\begin{lem}\label{lem:duality}
Strong duality holds for the primal problem
\begin{alignat*}{3}
p_\star \defeq\ \ &\makebox[\widthof{infimum}]{$\underset{\displaystyle s,\,P,\,\{\lambda_i\}}{\mathrm{infimum}}$} \quad && \mathrlap{s} \\
  &\text{subject to} \quad &&& s I       & \succeq \L(P) + \sum_{i\in\I} \lambda_i\,M_i \\
  &                        &&& P         & \succeq I \nonumber \\
  &                        &&& \lambda_i & \ge 0 \quad\forall i\in\I
\end{alignat*}
with $s\in\real$, $P\in\symmetric^n$, and $\lambda_i\in\real$ for all $i\in\I$, and its dual
\begin{alignat*}{3}
d_\star \defeq\ \ &\makebox[\widthof{maximum}]{$\underset{\displaystyle Q}{\mathrm{maximum}}$} \quad && \mathrlap{\trace\bigl(\L^*(Q)\bigr)} \\
  &\text{subject to} \quad &&& 0 & \preceq \L^*(Q) \\
  &                        &&& 0 & \le \trace(QM_i) \quad\forall i\in\I \\
  &                        &&& 0 & \preceq Q \\
  &                        &&& 1 & = \trace(Q)
\end{alignat*}
with $Q\in\symmetric^{n+m}$. In other words, $p_\star=d_\star$ and the dual optimum is attained if finite.
\end{lem}

\noindent\textbf{Proof.}
We can obtain a Slater point for the primal by taking $s$ sufficiently large, so strong duality holds. \qedhere

Using this primal-dual pair, we can then show that an alternative LMI is feasible when the spectral radius is equal to one; see~\cite{balakrishnan2003} for an overview on constructing strong alternatives for problems in control.

\begin{lem}\label{lem:Q}
Suppose $\rho(A,B,\M)=1$. Then there exists nonzero $Q\succeq 0$ such that $\L^*(Q)=0$ and $\trace(QM_i)\ge 0$ for all ${i\in\I}$.
\end{lem}

\noindent\textbf{Proof.}
Since $\rho(A,B,\M)=1$, the optimal value of the primal is $p_\star=0$. By strong duality, we have $d_\star=0$ and the dual optimum is attained; let $Q\in\symmetric^{n+m}$ denote the optimal solution. Since $\L^*(Q)\succeq 0$ and $\trace\bigl(\L^*(Q)\bigr)=0$, we have that $\L^*(Q)=0$. Also, $Q$ is nonzero since $\trace(Q)=1$. Therefore, $Q$ satisfies the LMI. \qedhere

Factoring $Q$ gives the matrices $X$ and $U$ in Lemma~\ref{lem:XUF}, and we obtain the matrix $F$ using the following lemma.

\begin{lem}\label{lem:rantzer}
Let $G$ and $H$ be real matrices of the same size. Then $G G^\tp = H H^\tp$ if and only if $G = H F$ for some orthogonal matrix $F$.
\end{lem}

\noindent\textbf{Proof.}
The case when $G$ and $H$ are complex is proved in~\cite[Lemma 3]{rantzer1996} using polar decompositions of $G$ and~$H$. Since the polar decomposition of real matrices is also real, the same proof may be used in the case when $G$ and $H$ are real matrices. \qedhere


\noindent\textbf{Proof of Lemma~\ref{lem:XUF}.}
Suppose $\rho(A,B,\M)=1$ and $B$ is full column rank. Then from Lemma~\ref{lem:Q}, there exists nonzero $Q\succeq 0$ such that $\L^*(Q)=0$ and $\trace(QM_i)\ge 0$ for all ${i\in\I}$. Denote a rank factorization of $Q$ by
\begin{align}\label{eq:Q}
Q = \bmat{X \\ U} \bmat{X \\ U}^\tp
\end{align}
with $X\in\real^{n\times d}$ and $U\in\real^{m\times d}$ where $d\defeq\rank(Q)$. Then applying Lemma~\ref{lem:rantzer} to the equation
\begin{align*}
0 = \L^*(Q) = (AX+BU) (AX+BU)^\tp - XX^\tp
\end{align*}
gives that $AX+BU=XF$ for some orthogonal matrix~$F$. Assume, by contradiction, that $X$ is zero. Then we have $B U U^\tp B^\tp=0$, which implies $U$ is zero since $B$ is full column rank. But then $Q$ is zero, which is a contradiction; therefore, $X$ is nonzero. Also, we have
\begin{align*}
0 \le \trace(QM_i) = \trace\Bigl(\bmat{X \\ U}^\tp\! M_i \bmat{X \\ U}\Bigr)
\end{align*}
for all $i\in\I$, so the tuple $(X,U,F)$ satisfies~\eqref{eq:XUF}. \qedhere


\begin{lem}\label{lem:orthogonal}
For any orthogonal matrix $F$, there is a subsequence of $\{F^k\}_{k=1}^\infty$ that converges to the identity.
\end{lem}

\noindent\textbf{Proof.}
The set of orthogonal matrices is compact \cite[Section 2.1]{horn2012} and therefore complete, so it has a Cauchy subsequence $\{F^{k_i}\}_{i=1}^\infty$ where the index sequence $\{k_i\}_{i=1}^\infty$ is monotonically increasing. Then for all $\epsilon>0$, there exists an integer $N$ such that
\begin{align*}
\epsilon > \|F^{k_i}-F^{k_j}\|
  = \|F^{k_j} (F^{k_i-k_j} - I)\|
  = \|F^{k_i-k_j} - I\|
\end{align*}
for all $i,j\ge N$. Since this holds for all $i\ge N$ and $k_i\to\infty$ as $i\to\infty$, the subsequence converges to the identity. \qedhere

\noindent\textbf{Proof of Theorem~\ref{thm:example}}. First, we have from~\eqref{eq:XUF1} that
\begin{align*}
A x_k + B u_k
  = (AX+BU) F^k v
  = X F^{k+1} v
  = x_{k+1},
\end{align*}
so the trajectory~\eqref{eq:trajectory} satisfies the dynamics~\eqref{eq:system}. Next, we show that the IQCs~\eqref{eq:soft} are also satisfied. Decomposing $F$ as in the technical condition and defining the matrices
\begin{align*}
H_i \defeq \bmat{X \\ U}^\tp\! M_i \bmat{X \\ U} \in \symmetric^d
\end{align*}
for each $i\in\I$, the IQC sum is
\begin{align*}
&\sum_{k=0}^{N-1} \bmat{x_k \\ u_k}^\tp\! M_i \bmat{x_k \\ u_k}
  = \sum_{k=0}^{N-1} v^\tp (F^k)^\tp H_i F^k v \\
  &\qquad = \sum_{k=0}^{N-1} v^\tp W (D^k)^* W^* H_i W D^k W^* v \\
  &\qquad = \sum_{j=1}^r \sum_{\ell=1}^r v^\tp W_j W_j^* H_i W_\ell W_\ell^* v \sum_{k=0}^{N-1} e^{\i k (\theta_\ell-\theta_j)}.
\end{align*}
Using the closed-form expression for the sum
\begin{align*}
\sum_{k=0}^{N-1} e^{\i k\theta} = \begin{cases} \dfrac{1-e^{\i N\theta}}{1-e^{\i\theta}} & \text{if } \theta \text{ is not a multiple of } 2\pi \\[3mm] N & \text{otherwise} \end{cases}
\end{align*}
and that all of the $\theta_j\in[0,2\pi)$ are distinct, we can lower bound the IQC sum by
\begin{align*}
&\sum_{k=0}^{N-1} \bmat{x_k \\ u_k}^\tp\! M_i \bmat{x_k \\ u_k}
  \ge N \sum_{j=1}^r v^\tp W_j W_j^* H_i W_j W_j^* v \\
  &\qquad\qquad - \sum_{\substack{j,\ell=1\\ j\neq\ell}}^r \bigl|v^\tp W_j W_j^* H_i W_\ell W_\ell^* v\bigr| \cdot \frac{2}{|1-e^{\i(\theta_\ell-\theta_j)}|}.
\end{align*}
The first term grows linearly with $N$ but is nonnegative since $v$ satisfies the technical condition, and the second term does not depend on $N$. Therefore, the IQC sum is uniformly lower bounded for all $N\ge 1$ and all $i\in\I$, so the IQCs are satisfied.

Finally, we show that the system is not asymptotically stable by proving the following inequalities:
\begin{align*}
0 < \|X v\| \le \limsup_{k\to\infty}\, \|x_k\| \le \|X\|\,\|v\|.
\end{align*}
The first inequality follows since $v$ is not in the null space of $X$, the second inequality since there is a subsequence of $\{\|x_k\|\}_{k=1}^\infty$ that converges to $\|X v\|$ from Lemma~\ref{lem:orthogonal}, and the final inequality from sub-multiplicativity of the matrix norm and orthogonality of~$F$. \qedhere

\section*{Appendix B --- Dynamic IQCs}\label{app:dynamic_IQC}

Our problem setup is different from that often used in the IQC literature (see~\cite{megretski1997,seiler2015} and the references therein) since we use static IQCs. We now show how to put an IQC with dynamics into our form.

To simplify notation, we consider the case of one IQC. Consider the linear time-invariant system
\begin{align*}
x_{k+1} &= A x_k + B u_k, \quad k\ge 0, \quad x_0\in\real^n \\
y_k     &= C x_k + D u_k
\end{align*}
with $m$ inputs and $m$ outputs subject to the frequency-domain IQC
\begin{align*}
\int_{-\pi}^\pi \bmat{\hat{y}(e^{\i\theta}) \\ \hat{u}(e^{\i\theta})}^*\! \Pi(e^{\i\theta}) \bmat{\hat{y}(e^{\i\theta}) \\ \hat{u}(e^{\i\theta})} \text{d}\theta \ge 0
\end{align*}
defined by the measurable Hermitian-valued function $\Pi : e^{\i\real}\to\complex^{(n+m)\times(n+m)}$ where $\hat{y}$ and $\hat{u}$ are the Fourier transforms of $y$ and $u$. By factoring $\Pi(z) = \Psi(z)^* M \Psi(z)$, where $\Psi(z)$ has the state-space representation
\begin{align*}
\psi_{k+1} &= A_\psi \psi_k + B_\psi^1 y_k + B_\psi^2 u_k \\
z_k        &= C_\psi \psi_k + D_\psi^1 y_k + D_\psi^2 u_k,
\end{align*}
Parseval's theorem can be used to show that~\eqref{eq:system} with
\begin{align*}
(A,B) \to \left(\bmat{A & 0 \\ B_\psi^1 C & A_\psi},\bmat{B \\ B_\psi^2 + B_\psi^1 D}\right)
\end{align*}
satisfies the IQC~\eqref{eq:soft} with
\begin{align*}
M \to \bmat{D_\psi^1 C & C_\psi & D_\psi^2 + D_\psi^1 D}^\tp M \bmat{\star},
\end{align*}
where $\star$ denotes the corresponding symmetric part and the state is now the combined state of the original system with that of the dynamic IQC, i.e., $x_k \to \left[\begin{smallmatrix} x_k \\ \psi_k\end{smallmatrix}\right]$. Instead of this formulation, however, we study the system~\eqref{eq:system} subject to the IQCs~\eqref{eq:soft} to simplify the notation.

\bibliographystyle{abbrv}
{\small \bibliography{counterexamples}}

\end{document}